\documentclass[prl,twocolumn,showpacs,superscriptaddress,amsmath]{revtex4}
\usepackage{graphicx}
\usepackage{amssymb}
\usepackage{citesort}

\begin{document}
\title{Absence of superconductivity in the 
half-filled band 
Hubbard model on the anisotropic triangular lattice}
\author{R.T. Clay}
\affiliation{Department of Physics and Astronomy and HPC$^2$ Center for 
Computational Sciences, Mississippi State University, Mississippi State MS 39762}
\author{H. Li}
\affiliation{ Department of Physics, University of Arizona
Tucson, AZ 85721}
\author{S. Mazumdar}
\affiliation{ Department of Physics, University of Arizona
Tucson, AZ 85721}
\date{\today}
\begin{abstract}
We report exact calculations of magnetic and superconducting pair-pair
correlations for the half-filled band Hubbard model on an anisotropic
triangular lattice. Our results for the magnetic phases are similar to
those obtained with other techniques.  The superconducting pair-pair
correlations at distances beyond nearest neighbor decrease
monotonically with increasing Hubbard interaction $U$ for all
anisotropy, indicating the absence of frustration-driven
superconductivity within the model.
\end{abstract}

\maketitle

The possibility of superconductivity (SC) in the doped square-lattice
Hubbard model continues to be discussed in the context of cuprate
superconductors. Theoretical results for the strength of
superconducting pair correlations in this model are controversial, and
arguments for \cite{Maier05a,Plekhanov05a} and against
\cite{Zhang97b,Aimi07a} long range order (LRO) both exist in the
literature.  Yet another class of superconductors which have been
discussed using the Hubbard model are the organic charge-transfer
solids (CTS) \cite{Ishiguro}, which are known to have strong
electron-electron interactions \cite{strongcorr}.  SC in the CTS is
reached from proximate insulating states not by doping, but by
application of pressure at constant carrier density.  It is therefore
natural to expect that a simple change of parameters within a model
appropriate for the insulating states of the CTS can give SC. The
proponents of spin fluctuation and resonating valence bond models for
SC in the CTS suggest that precisely such an
insulator-to-superconductor transition can occur within the
$\frac{1}{2}$-filled band triangular lattice Hubbard model with
increasing frustration
\cite{Vojta98a,Schmalian98a,Kino98a,Kondo98a,Powell,Baskaran03a,Gan,Sahebsara06a,Watanabe06a,Kyung06a}.
\begin{figure}[tb]
\centerline{\resizebox{3.2in}{!}{\includegraphics{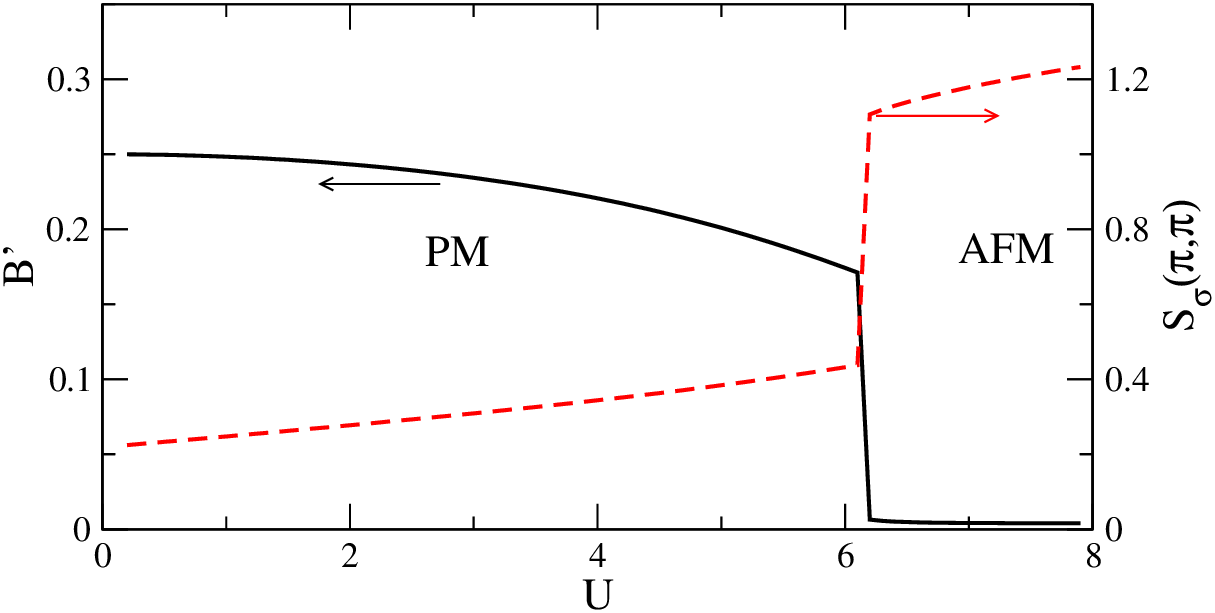}}}
\caption{(color online) Correlation functions as a function of $U$ for
$t^\prime$=0.5.  Solid line: bond order $B^\prime$ along the diagonal
bonds.  Dashed line: $S_\sigma(\pi,\pi)$.  Both are discontinuous at
the AFM--PM transition at $U=6.1$. Lines are guides to the eye based
on calculations performed on a grid of size $\Delta U=0.1$.}
\label{correlations}
\end{figure}

The frustrated Hubbard model has been applied most commonly to the CTS
belonging to the the $\kappa$-(BEDT-TTF)$_2$X ($\kappa$-(ET)$_2$X)
family with the highest critical temperature. In these materials, the
ET molecules form anisotropic triangular lattices of dimer pairs
\cite{Ishiguro}.  While $\kappa$-(ET)$_2$X is formally
$\frac{3}{4}$-electron filled ($\frac{1}{4}$-hole filled), considering
the ET dimers as unit cells the effective bandfilling becomes
$\frac{1}{2}$.  It has been suggested that pressure modifies the
interdimer electron hoppings and changes the degree of frustration
\cite{Kino95a}. Experimentally, both an antiferromagnetic (AFM) state
\cite{Miyagawa95a} in systems with relatively large anisotropy, and a
spin liquid state \cite{Shimizu03a} for nearly isotropic triangular
lattices, have been found as well as SC.  Theoretically, strong enough
frustration within the Hubbard model destroys the AFM order, giving a
paramagnetic metal (PM).  Within theories of SC mediated by AFM
fluctuations, the SC occurs in a narrow region sandwiched between the
AFM and PM.
\cite{Vojta98a,Schmalian98a,Kino98a,Kondo98a,Powell,Baskaran03a,Gan,Sahebsara06a,Watanabe06a,Kyung06a}.

We consider the Hamiltonian,
\begin{eqnarray}
H &=& - t\sum_{\langle ij \rangle,\sigma}(c_{i,\sigma}^\dagger c_{j,\sigma}+ H.c.) 
-t^{\prime}\sum_{[kl],\sigma}(c_{k,\sigma}^\dagger c_{l,\sigma}+ H.c.) 
\nonumber  \\
&+&U \sum_{i} n_{i,\uparrow} n_{i,\downarrow} 
\label{ham}
\end{eqnarray}
In Eq.~\ref{ham} $c^{\dagger}_{i,\sigma}$ creates an electron with
spin $\sigma$ ($\uparrow$, $\downarrow$) on site $i$,
$n_{i,\sigma}=c^{\dagger}_{i,\sigma}c_{i,\sigma}$.  $U$ is the on-site
Hubbard interaction. The lattice structure is the conventional square
lattice (hopping integrals $t$) with additional hopping integrals
$t^\prime$ across the $x+y$ diagonals for a total of three bonds per
site. The ratio $t^\prime/t$ is 0 for the square lattice and 1 for the
isotropic triangular lattice.  In what follows, we express all
quantities in units of $t$.

Accurate solutions to Eq.~\ref{ham} are difficult to reach using
existing analytical approaches. Many of the techniques that find SC
are based on mean-field embedded cluster methods that use extremely
small cluster sizes, including some as small as 4 or 8 sites
\cite{Kyung06a,Sahebsara06a}.  Eq.~\ref{ham} and a variant with two
diagonal hoppings have recently been investigated within a numerically
more precise Path Integral Renormalization Group (PIRG) technique
\cite{Imada}.  The phase diagrams found by these authors contain PM,
AFM, and nonmagnetic insulator (NMI) phases only and suggest
implicitly the absence of SC along the AFM--PM boundary. The authors,
however, did not calculate the SC pair-pair correlations. Recent exact
diagonalization calculations also investigated the AFM--PM phase
boundary, but did not calculate pair-pair correlations
\cite{Koretsune07a}. Specifically because of the difficulty of
calculating SC pair-pair correlations, and the prediction of SC
occurring only over a narrow region of the phase diagram, it is
essential that SC correlations are calculated explicitly with high
precision over the complete phase space.

We report precisely such calculations within Eq.~\ref{ham} via exact
diagonalization in the present Letter.  We focus here solely on
$\frac{1}{2}$-filling, numerical convergence for which is reached much
faster than in doped systems.  Although exact diagonalizations cannot
necessarily establish the {\it occurrence} of SC, they can test
precisely the {\it necessary conditions} for SC.  First, if the model
does indeed have a SC ground state with pair-pair correlations
exhibiting LRO in the thermodynamic limit, finite clusters should
necessarily exhibit short range order (SRO) in the pair-pair
correlations.  Second, if SC driven by the Hubbard $U$ interaction is
present, then there must exist some parameter region where $U$
enhances the pair-pair correlations relative to $U=0$.  Note that for
the one strongly-correlated model where it is known that SC exists,
the negative-$U$ Hubbard model, both the above conditions are met
\cite{Scalettar89a}.  We show here that neither of the above criteria
are met within Eq.~\ref{ham} for repulsive $U$ at
$\frac{1}{2}$-filling. For all $U$ and anisotropy, the SC pair
correlations become monotonically weaker with increasing $U$.

We solve exactly the ground state of Eq.~\ref{ham} on a periodic
4$\times$4 lattice.  We calculate the spin structure factor
$S_\sigma(\bf{q})$ and the bond order $B_{ij}$ between sites $i$ and
$j$,
\begin{eqnarray}
\hspace*{-0.1in}S_\sigma({\bf q})&=&\frac{1}{N} \sum_{{\bf r}{\bf r^\prime}} e^{i{\bf q}
 \cdot ({\bf r}
-{\bf r^\prime})} \langle (n_{\uparrow{\bf r}}-n_{\downarrow{\bf r}})
(n_{\uparrow{\bf r^\prime}}-n_{\downarrow{\bf r^\prime}})
\rangle \label{eqn-sfac} \\
B_{ij}&=& \sum_{\sigma} \langle c^{\dagger}_{i,\sigma} c_{j,\sigma} + H.c.
 \rangle  \label{Bond-order} 
\end{eqnarray}
In Eq.~\ref{eqn-sfac} $N$ is the number of lattice sites.  We 
define the standard pair-creation operator $\Delta^\dagger_i$ 
\begin{equation}
\Delta^\dagger_i= \frac{1}{\sqrt{2}}\sum_{\nu} g(\nu) 
(c^\dagger_{i,\uparrow}c^\dagger_{i+{\bf \nu},\downarrow}
- c^\dagger_{i,\downarrow}c^\dagger_{i+{\bf \nu},\uparrow})
\label{pair}
\end{equation}
In Eq.~\ref{pair} the phase factor $g(\nu)$ determines the symmetry
of the superconducting pairs.  For $s$-wave pairing $g(\nu)$ is +1 for
the four nearest-neighbor sites $i+\hat{x}$, $i+\hat{y}$,
$i-\hat{x}$ and $i-\hat{y}$; for $d_{x^2-y^2}$ pairing $g(\nu)$ alternates as
+1,-1,+1,-1 for the same four sites; and for $d_{xy}$ pairing $g(\nu)$
alternates sign over the four sites $i+\hat{x}+\hat{y}$,
$i+\hat{x}-\hat{y}$, $i-\hat{x}-\hat{y}$, $i-\hat{x}+\hat{y}$.  We
will calculate the pair-pair correlation function as a function of
distance, $P(r)=\langle \Delta^\dagger_i\Delta_{i+\vec{r}} \rangle$.

\begin{figure}[tb]
\centerline{\resizebox{3.2in}{!}{\includegraphics{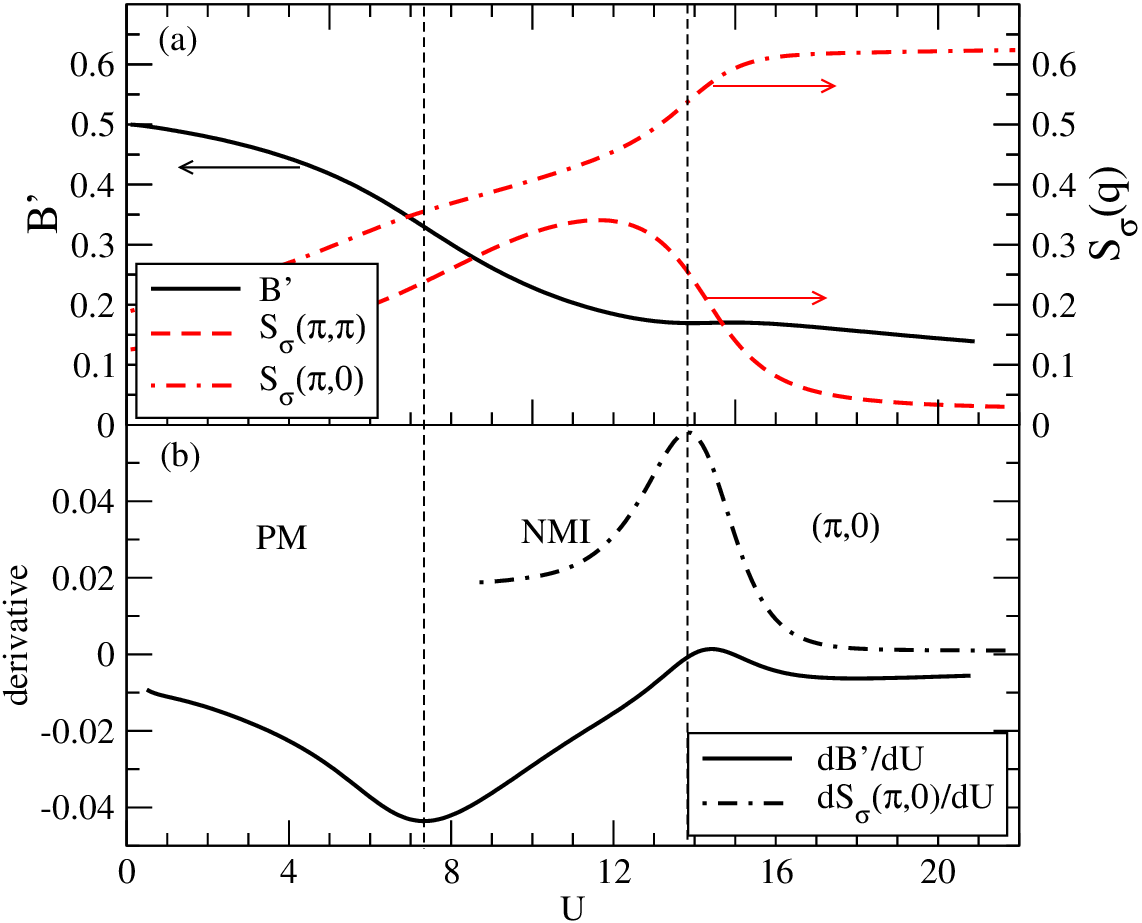}}}
\caption{(color online) Correlation functions as a function of $U$ for
$t^\prime$=1.1. (a) Diagonal bond order $B^\prime$ (solid line),
$S_\sigma(\pi,\pi)$ (dashed line), and $S_\sigma(\pi,0)$ (dot-dashed
line).  (b) Derivative of $B^\prime$ and $S_\sigma(\pi,0)$ with
respect to $U$. Vertical lines show estimated phase boundaries between
PM, NMI, and ($\pi$,0) phases.  Lines are guides to the eye and based
on calculations performed on a grid of size $\Delta U=0.1$;
derivatives calculated with a three-point centered difference
formula.}
\label{correlations-1.1}
\end{figure}
We first discuss the non-SC phases within the model.  In the limit
$t^\prime=0$, the ground state of Eq.~\ref{ham} has AFM LRO for any
nonzero $U$. This is usually determined from plots of $S_\sigma({\bf
q})$ versus ${\bf q}$, which exhibit strong peaks at ${\bf
q}=(\pi,\pi)$.  For $t^\prime \neq 0$, the ground state is AFM only
when $U$ exceeds a critical value $U_c(t^\prime)$ and remains a PM for
smaller $U$.  We determine $U_c$ from calculations of
$S_\sigma(\pi,\pi)$, the bond order corresponding to the diagonal
$t^\prime$ bonds ($B^\prime$), and the ground state expectation value
of the double occupancy $d=\langle n_{i,\uparrow} n_{i,\downarrow}
\rangle$.  In Fig.~\ref{correlations} we plot $S_\sigma(\pi,\pi)$ and
$B^\prime$ for $t^\prime=0.5$.  It is in this region of $t^\prime$
that the $d_{x^2-y^2}$ SC phase is claimed to be broadest
\cite{Kyung06a}.  $U_c$ is identified from a discontinuous decrease in
$B^\prime$ accompanied by a simultaneous increase in
$S_\sigma(\pi,\pi)$.  The double occupancy $d$ also decreases
discontinuously {\it at the same $U_c$} (not shown).  The large
decrease of $B^\prime$ and $d$ indicate loss of electron mobility
associated with a metal-insulator (M-I) transition, while the sharp
increase in $S_\sigma(\pi,\pi)$ indicates the AFM nature of the
insulator.  The {\it simultaneous} discontinuity of all observables at
$U_c$ is consistent with a first-order (discontinuous) quantum phase
transition.  The discontinuity continues to exist for $t^\prime<1$.
The size of the discontinuity in $B^\prime$ decreases as $t^\prime$
approaches 1; this is likely related to the appearance of the NMI
phase as well as the change in the magnetic ordering at $t^\prime=1$
we discuss below.  We also performed calculations (not shown here) on
a 12 site cluster, chosen so that $(\pi,\pi)$ AFM order is not
frustrated at $t^\prime=0$, and found comparable values of $U_c$.

The magnetic ordering is qualitatively different for $t^\prime>1$,
where $S_\sigma({\bf q})$ peaks at ${\bf q}=(\pi,0)$ instead of
(${\pi,\pi}$) \cite{Imada}, with the amplitude of the peak increasing
with $U$.  This transition is expected from the known results for the
antiferromagnetic Heisenberg spin Hamiltonian with nearest ($J_1$) and
next-nearest ($J_2$) neighbor exchange interactions (two diagonal
bonds per site) \cite{Nevidomskyy08a}.  Here $J_2/J_1=0.5$ is a
quantum critical point, with $\bf{q}=(\pi,\pi)$ dominating for
$J_2/J_1<0.5$, and ${\bf q}=(\pi,0)$ for $J_2/J_1>0.5$
\cite{Kotov00a}. This value of $J_2$ corresponds to
$t^\prime=1/\sqrt{2}$ within the $U\rightarrow\infty$ limit of the
Hubbard model with two diagonal bonds, and this is where the
transition to the ($\pi$,$0$) ordering is found
\cite{Nevidomskyy08a}. From perturbation theory, the quantum critical
$J_2/J_1$ shifts to 1 when the number of diagonal bonds per site is
one instead of two, corresponding to $t^\prime=1$ within the large $U$
Hubbard model. This is exactly where we have found the transition in
our numerical work.

In Fig.~\ref{correlations-1.1}(a) we show $B^\prime$ and spin
structure factor data for $t^\prime=1.1$ as a function of $U$.  Unlike
the discontinuous transition between the PM and AFM phases in
Fig.~\ref{correlations}, the correlation functions change continuously
now.  Importantly, the steepest decrease in $B^\prime$ and the changes
in $S_\sigma({\bf q})$, for both ${\bf q}=(\pi,\pi)$ and ${\bf
q}=(\pi,0)$, occur at different $U$, suggesting the existence of a
third phase between the PM and the $(\pi,0)$ magnetic order.  In
Fig.~\ref{correlations-1.1}(b) we show the derivatives of $B^\prime$
and $S_\sigma(\pi,0)$ with respect to $U$. At $U=7.3$ a minimum occurs
in $dB^\prime/dU$, which is accompanied by a minimum in the derivative
of the double occupancy $d$ at nearly the same value of $U$ (not
shown).  At $U=13.8$ similar maxima occur in derivatives of $B^\prime$
and $S_\sigma(\pi,0)$. We therefore identify three distinct phases: a
PM phase for $U\leq 7.3$, a nonmagnetic (NM) phase for $7.3 < U
\leq13.8$, and the magnetically ordered $(\pi,0)$ phase for $U>13.8$.
The transport behavior within the NM phase is difficult to determine
from our finite-cluster calculations. The continuous decrease of
$B^\prime$ and $d$ in the intermediate U region is, however,
consistent with an insulating character of this phase. We therefore
identify the intermediate U-phase as the previously discussed NMI
phase \cite{Imada}.
\begin{figure}[tb]
\centerline{\resizebox{3.2in}{!}{\includegraphics{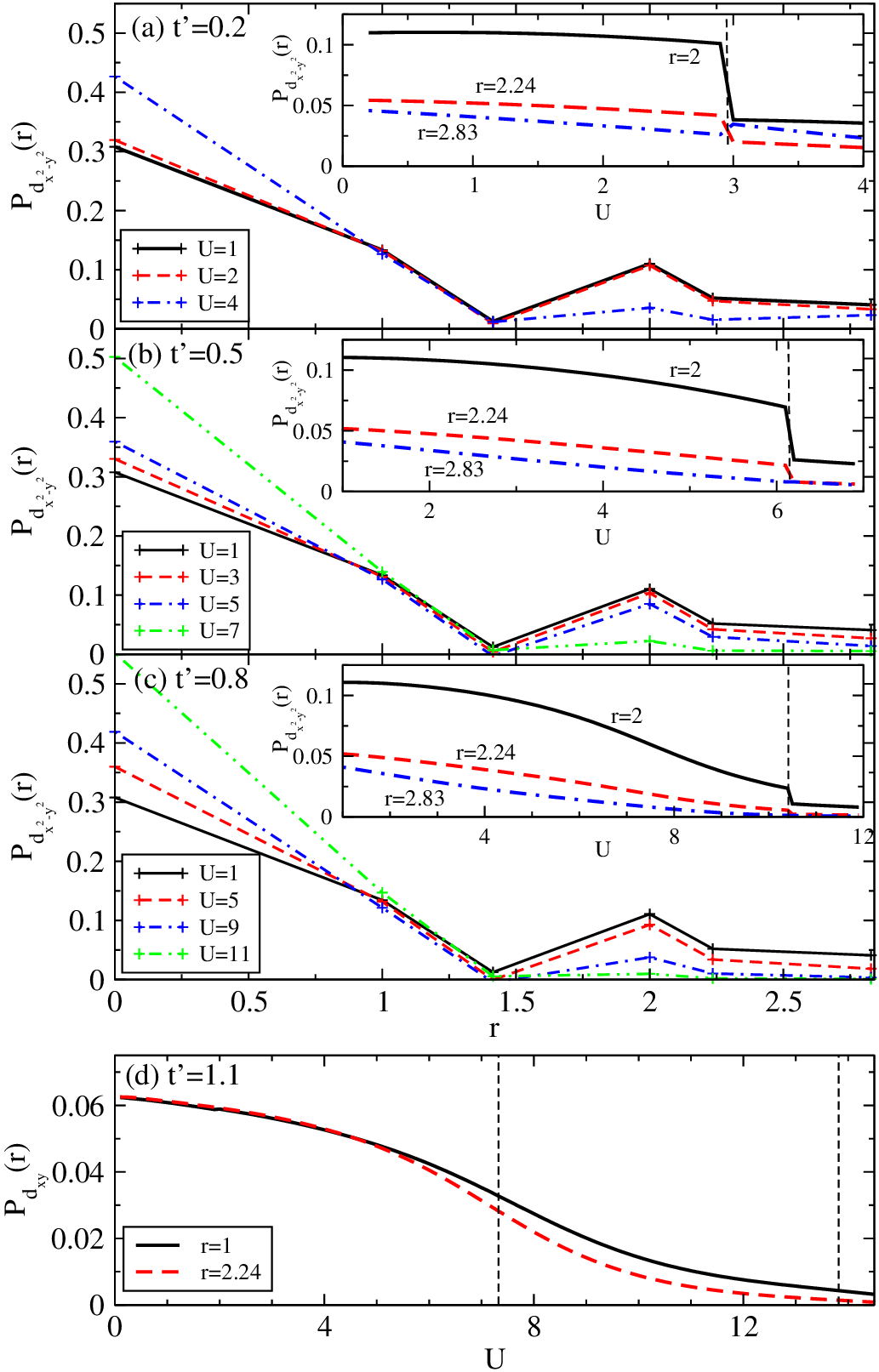}}}
\caption{(color online) (a)-(c) $d_{x^2-y^2}$ pair-pair correlation
functions as a function of $r$ for $t^\prime$=0.2, 0.5 and 0.8,
respectively.  Insets show $U$-dependence of $d_{x^2-y^2}$ pair-pair
correlation functions for three different $r$.  The PM--AFM transition
is indicated by vertical lines.  (d) U-dependence of the $d_{xy}$
pair-pair correlations at two different $r$ for for $t^\prime=1.1$.
Vertical lines mark PM--NMI and NMI--($\pi,0$) transitions.  Lines are
guides to the eye and based on calculations performed on a grid of
size $\Delta U=0.1$.}
\label{pr}
\end{figure}

Mean field calculations finding SC generally agree that the pairing is
of $d_{x^2-y^2}$ symmetry for $0<t^\prime<1$.  In Fig.~\ref{pr}(a) -
(c) we show the $d_{x^2-y^2}$ singlet pair-pair correlation function
$P_{d_{x^2-y^2}}(r)$ as a function of distance $r$ and Hubbard
interaction $U$, for $t^\prime$ = 0.2, 0.5 and 0.8, respectively.  It
is for the intermediate $t^\prime=0.5$ region that SC correlations
have been claimed to be the strongest. The insets show the behavior of
$P_{d_{x^2-y^2}}(r)$ as a function of $U$ for $r=2, \sqrt{5}$ and
$2\sqrt{2}$.  $P_{d_{x^2-y^2}}(r)$ at the two larger values of $r$
that we have chosen are least affected by the finite lattice size,
since there is no overlap between pairs on the 4$\times$4 lattice for
these $r$.

Several features in the pair-pair correlations are immediately
apparent.  First, the pair-pair correlations decrease rapidly with
distance, the values at the largest $r$ being more than an order of
magnitude smaller than those at $r=0$. Second, like the other
correlation functions (Fig.~\ref{correlations}), $P_{d_{x^2-y^2}}(r)$
changes discontinuously at the PM--AFM transition.  Finally, for
$r>1$, the pair-pair correlations are strongest at $U=0$ and decrease
{\it monotonically} with increasing $U$.  Pair-pair correlations at
$r=0$ are trivially enhanced with increasing $U$.  Nearest-neighbor
correlations ($r=1$) decrease with $U$ for $U<U_c$ for all $t^\prime$
but increase for $U>U_c$ for large enough $t^\prime$. This increase is
associated with the transition to AFM and occurs {\it only on the
insulating side of the PM--AFM boundary.}  Furthermore, the small
increase in the $r=1$ correlations is accompanied by a large
simultaneous decrease in the long-range component, consistent with the
entry into an insulating state (clearly seen in the $U=4$ data in
Fig.~\ref{pr}(a) and the $U=7$ data in Fig.~\ref{pr}(b).)  This
behavior of the short-range correlations may be one reason embedded
cluster methods using small cluster sizes find SC.
\begin{figure}[tb]
\centerline{\resizebox{3.2in}{!}{\includegraphics{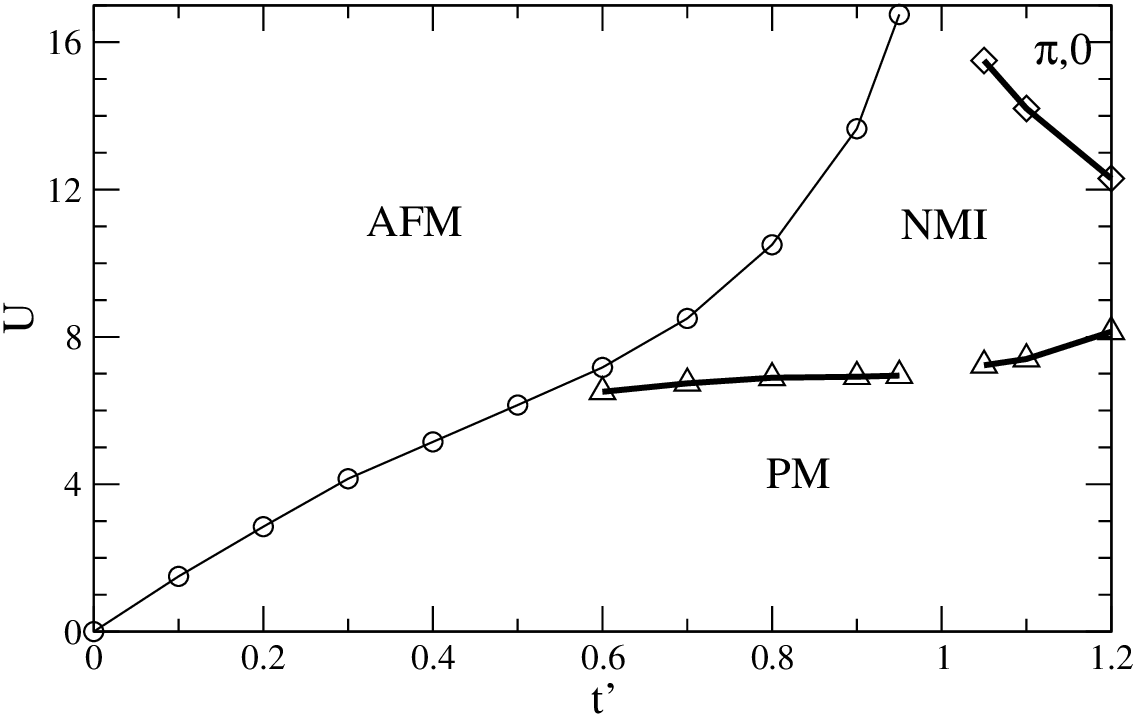}}}
\caption{Phase diagram as a function of $t^\prime$ and $U$.  Lines are
guides to the eye. The PM--AFM boundary is located to within $\Delta
U=0.1$ using the discontinuity of correlation functions. The PM--NMI
and NMI--$(\pi,0)$ phase boundaries are drawn from the inflection
point in the correlation functions (see text.)  }
\label{phasediag}
\end{figure}

Based on the ${\bf q}=(\pi,0)$ magnetically ordering for $t^\prime>1$,
it has been proposed that the SC correlations have pairing symmetry
$d_{xy}$ or $s+d_{xy}$ \cite{Li03a,Gan,Powell}. In Fig.~\ref{pr}(d) we
show the $U$-dependence for $d_{xy}$ pair-pair correlations at
$t^\prime=1.1$, for the two separations where the pairs are
nonoverlapping.  Exactly as for $t^\prime<1$, we find no region of $U$
where the ``long-range'' $d_{xy}$ pair-pair correlations are enhanced
compared to their $U=0$ values.  We have performed similar
calculations for $s$ and $s+d_{xy}$ pairing symmetries, and find no
enhancement for these either.

We give the phase diagram from our calculations in
Fig.~\ref{phasediag}. We have not plotted data for $t^\prime$ exactly
1 due to problems associated with degeneracy at this point. We have
found that the NMI phase not only exists for $t^\prime>1$, but enters
the phase diagram at approximately $t^\prime\agt 0.6$; for all
$t^\prime>0.5$ inflection points as in Fig.~\ref{correlations-1.1}
appear in $B^\prime$ and $d$.  Our phase diagram for the insulating
states and the boundaries between them are in good agreement with
other numerical results
\cite{Imada,Kyung06a,Koretsune07a,Sahebsara08a,Nevidomskyy08a}.
We find, for example, that the minimum $U$ necessary for the NMI phase
is $U\sim 7$. This is very close to the boundary of $U\sim 6.7$ at
$t^\prime=1$ found recently using a variational cluster approximation
\cite{Sahebsara08a}.  The data in Fig.~\ref{correlations-1.1} suggest
that the PM--NMI transition is continuous \cite{Imada}, but
confirmation of this would require finite-size scaling.

Our most important result is that SC pair-pair correlations do not
exhibit even SRO within Eq.~1, in any channel and for any range of
parameters. Theoretical models for SC in the CTS should either include
electron-phonon interactions, or should be based on a different
electronic model. We have recently suggested a correlated-electron
theory of SC in the CTS that emphasizes their $\frac{1}{4}$-filled
band nature, and the transition to the SC state from a density wave
composed of local singlets \cite{Mazumdar08a}. Such a theory can
explain naturally the recently observed transition from a valence-bond
solid to the SC state \cite{Shimizu07a}

This work was supported by the Department of Energy grant DE-FG02-06ER46315.
Work at Arizona was also partially supported by NSF-DMR-0705163.

\end{document}